\documentclass[a4paper,12pt]{article}

\usepackage[affil-it]{authblk}
\usepackage{amsmath}
\usepackage{amssymb}
\usepackage{graphicx}
\usepackage{subfig}
\usepackage{bbm}
\usepackage{epsfig}
\usepackage{yfonts}

\begin{document}

\title{Non-equilibrium steady state in the hydro regime}
\author{Razieh Pourhasan\footnote{e-mail: {\tt
razieh@hi.is}}}

\affil{Science Institute, University of Iceland, Dunhaga 5, 107
Reykjavik, Iceland}

\maketitle

\begin{abstract}
We study the existence and properties of the non-equilibrium steady state which arises by putting two copies of systems at different temperatures into a thermal contact. We solve the problem for the relativistic systems that are described by the energy-momentum of a perfect hydro with general equation of state (EOS). In particular, we examine several simple examples: a hydro with a linear EOS, a holographic CFT perturbed by a relevant operator and a barotropic fluid, i.e., $P=P(\mathcal{E})$. Our studies suggest that the formation of steady state is a universal result of the hydro regime regardless of the kind of fluid.

\end{abstract}
\newpage

\section{Introduction}

Although the study of equilibrium states has been the focus of most research in many body systems and condensed matter physics, much of the interesting phenomena around us are far from equilibrium. However, thermodynamic study of non-equilibrium states are less advanced since the dynamical equations governing thermodynamic variables are highly non-linear differential equations. Unlike the thermodynamic equilibrium where the processes are reversible and time independent, for a system in a non-equilibrium state processes are generally irreversible and time dependant. That's what makes the study of systems far from equilibrium so difficult. Nevertheless, waiting long enough, most non-equilibrium systems tend to approach to the state of thermodynamic equilibrium unless there is a continuous flow of matter or energy to or from the system. 

Yet an interesting subclass of non-equilibrium systems is the \textit{steady state} which is reached by a system subject to a constant driving force. In a non-equilibrium steady state (NESS) there is no time variation, i.e., all thermodynamic properties are time independent, although the entropy production is non-zero and the system is recognised by the presence of fluxes. The fluxes are the flows of currents of the conserved variables driven by the gradient of their corresponding thermodynamic conjugate, e.g., the energy current is driven by the temperature gradient, and so on. Since the dynamics of such a system is dissipative, in order to maintain the steady state one needs the injection of energy at one boundary and subtraction at the other. One simple example of such NESS is a system confined between two heat baths (reservoirs) at different but constant temperature, transporting heat from one to another by the energy current which is generally proportional to the temperature gradient. 

However in a recent studies of one dimensional quantum critical systems described by conformal field theory (CFT), Bernard and Doyon \cite{Bernard:2013} found an interesting steady state with a nontrivial energy current choosing a different framework; Instead of putting a system in contact with external reservoirs, they picked two copies of some large quantum systems at different thermal equilibrium, e.g., different temperatures, and then glued them together at a contact point which result in the energy transfer from one to another. After long enough time, a steady state would be established and parts of each system far from the contact point would effectively behave like heat baths. Another interesting point about their set-up is that although the temperature profile is flat, there exists a non-zero energy current. That is, unlike the usual case for steady states, the energy current  is \textit{not} simply a function of temperature gradient. Moreover, the universal character of the resulting steady state is noticeable; the energy transfer only depends on the universal constants and the temperatures of the initial copies they start with. 

Motivated by these results, the existence of such a steady state and its universality has been investigated for CFT in higher dimensions \cite{Chang:2013gba, Bhaseen:2013ypa}. Therefore, one legitimate question might be whether this is a particular feature of CFT, or one may get similar answer starting from initial systems with more general equation of state. In this manuscript we are trying to explore this question relying on the holographic insight given in \cite{Bhaseen:2013ypa};\footnote{We prefer to follow the method introduced in \cite{Bhaseen:2013ypa} as opposed to the ansatz given in \cite{Chang:2013gba}. The former,  due to holographic analysis, has less parameters and is easier to deal with particularly if one interested in studying a more general equation of state like barotropic case that we investigate in this paper.} there it has been shown that the non-equilibrium steady state on the conformal boundary is in fact dual to the Lorentz boosted black brane in the bulk. According to the AdS/CFT correspondence, the stress tensor of the steady state could be obtained from the metric of the dual gravity theory in the bulk. Therefore, the steady state is completely specified if one obtains its temperature and boost velocity, i.e., $\{T,\,v\}$, in terms of the temperatures of the initial systems at the instant of contact, denoted by $T_l$ and $T_r$ for one system on the left and the other on the right. In fact, at sufficiently large scales two initial systems look like asymptotic reservoirs with the steady state as the intermediate state interpolating between two reservoirs on the left and right. Then in this framework, the problem is similar to a Riemann problem with initial boundary condition $T_l$ for $x<0$ and $T_r$ for $x>0$, where a possible solution would be two shock waves emitting at the point of contact and moving to the opposite directions. Note that the development of the steady state in this set-up also requires that two shock waves propagate in the opposite directions without splitting and decomposition, for if this happens the composite waves could lead to cascades and finally thermalizing the system. 


Riemann problem for classical and relativistic hydrodynamics with various equation of states is widely studied in the literature, see for example \cite{Lax:1957,Lax:1972, Menikoff:1989ka, Smoller, FLM:357096}. We follow similar approach of solving Riemann problem to study NESS while considering to have a relativistic perfect fluids for the left and right hydro before bringing them into a thermal contact. We solve the problem assuming a general equation of state, e.g., barotropic fluid. In general, we conclude that the formation of steady state after thermal contact has nothing to do with conformal field theory or even with integrable models; it is a universal result in the hydro regime for just about any kind of fluid.

Therefore, the outline of this paper is as follows: in section \ref{shocks} we briefly review shock wave solutions to the Riemann problem as well as Rankine–-Hugoniot jump condition which are essential in our calculations of specifying the steady state. Then in section \ref{examples}, starting with the perfect fluid assumption for the initial systems at the instant of contact, we investigate the steady state properties for the systems with more general equation of state (EOS) than the conformal fluid. In particular, we study three examples:  systems with linear equation of state, the QFT model described in \cite{Buchel:2013lla} as a small deviation from CFT, and finally systems with barotropic equation of state where pressure is only a function of energy density however we restrict our calculation to the case where the temperature difference between two systems are very small. We close the paper by some discussion about limitation on the stability and the existence of shock solutions and also give a brief conclusion in section \ref{conclusion}.

\section{Shock waves and jump condition}\label{shocks}

Consider a relativistic fluid described by the energy density $\mathcal{E}$ and
the pressure $P$, and then introduce a small perturbations to the
pressure and density of the system. It is straightforward to show
that these perturbations, the so-called ``{\it sound waves}'',
propagate through the fluid with the velocity $c_s$ given by
\begin{equation}
c_s^2\equiv\frac{dP}{d \mathcal{E}}\,,\label{soundspeed}
\end{equation}
known as the ``{\it speed of sound}'' in literature. If we assume
that the sound waves have infinitesimal amplitude, i.e., the
perturbations are very small, then the speed of sound is nearly
uniform throughout the fluid. However, in general the speed of
sound is a function of density which result in the crest of wave
to move faster than the trough. When the crest overtakes the
trough a {\it shock wave} can form due to the steepening of wave.
A shock front is a surface that marks a sudden jump in the density
and pressure of the fluid. Although this is a way that shocks
could form, in general, shock waves are characterized by a rapid,
discontinuous change in the density and pressure of the system.
The shock is bounded into an infinitesimal region where the fluid
properties, such as density and pressure, immediately before and
after being shocked are linked by the jump conditions. These
conditions, usually referred to as the \textit{Rankine–-Hugoniot
jump conditions}, are derived from the conservation laws.

Consider a conserved quantity $q(x,t)$ in (1+1)-dimensions\footnote{The hyperbolic conservation laws originally studied in \cite{Lax:1957,Lax:1972} for a theory in (1+1)-dimensions. However, one can follow a similar approach for a theory in higher dimensions by imposing a symmetry so that nothing happens in the remaining transverse directions.}
satisfying the hyperbolic conservation law
\cite{Lax:1957,Lax:1972}, i.e.,
\begin{equation}
\partial_t q+\partial_x f(q)=0\,,\label{hyper}
\end{equation}
where $f$ represents the flux of $q$. Now assume that there exists
a solution $q$ which has a discontinuity, i.e., a shock, along the
curve $x=\xi(t)$; this is known as the {\it weak solution} of the
PDE (\ref{hyper}) \cite{Bressan}. Then choose the interval
$\xi_-\leq x\leq \xi_+$ such that it intersects with $x=\xi(t)$ at
time $t$. Integrating equation (\ref{hyper}) over this interval,
yields
\begin{eqnarray}
\frac{d}{dt}\left(\int_{\xi_-}^{\xi(t)}q(x,t)dx+\int_{\xi(t)}^{\xi_+}q(x,t)dx\right)+\int_{\xi_-}^{\xi_+}\partial_xf(q(x,t))dx=0\,.\label{integral}
\end{eqnarray}
If we apply Leibnitz rule to (\ref{integral}) using the fact that
the conservation equation (\ref{hyper}) is satisfied on either
side of the discontinuity, we obtain
\begin{equation}
\frac{d\xi(t)}{dt}(q_l - q_r)=f_l-f_r\,,\label{jump}
\end{equation}
where $q_l$ and $q_r$ denote the values of $q$ on the left and
right sides of $x = \xi(t)$, respectively, and
$$f_l=f(q_l)\,,\qquad f_r=f(q_r)\,.$$ Equation (\ref{jump}) is called
the \textit{Rankine-Hugoniot jump condition} and is usually
written in shorthand notation as
\begin{equation}
u_s[q]=[f]\,,
\end{equation}
where $[q]\equiv q_l - q_r$ and $[f]\equiv f_l-f_r$ are the jumps
across the discontinuity (or shock) and $u_s\equiv d\xi(t)/dt$ is
referred to as the \textit{shock speed}.

In more physical systems there are more than one conservation law
and the dynamical behavior of the fluid is governed by a system of
hyperbolic conservation laws as
\begin{equation}
\partial_t Q + \partial_x F=0\,,\label{hypersys}
\end{equation}
where $Q=(q_1,\ldots,q_n)$ and $F$ is a function of
$q_1,\ldots,q_n$. Therefore, one would derive $n$ jump conditions
corresponding to each equation in (\ref{hypersys}) as
\begin{equation}
u_s[Q]=[F]\,.
\end{equation}
However, to obtain a full description of the fluid, we also need
to take into account the equation of state (EOS) which relates
different thermodynamic variables of the system. Although the
conservation laws and the jump conditions are valid for any EOS,
the deterministic role of EOS on the nature of the shock waves and
their propagations has been widely considered in literature, see
for example \cite{Bethe, Zeldovich, Thompson}.

Therefore, a key concern in studying the fluid dynamics is the
existence and uniqueness of the shock solutions in a given fluid,
the so-called \textit{Riemann problem}. In general, the Riemann
problem is the initial value problem for a system of conservation
laws where the initial data are a pair of constant states separated
by a jump discontinuity at $x=0$, i.e.,
\begin{equation}
Q_0(x)=\left\{ \begin{array}{ll}
         Q_l \quad& \mbox{if $x < 0$},\\
        Q_r \quad& \mbox{if $x > 0$}.\end{array} \right.
\end{equation}

In the next section, we study steady state solutions for the
perfect fluid with various EOSs and examine the constraints to
have the solution to the Riemann problem.

\section{Perfect fluid and steady state solution}\label{examples}

Consider two isolated semi-infinite systems each described by a
relativistic perfect fluid, i.e.,
\begin{equation}
T^{\mu\nu}=(P+\mathcal{E})u^{\mu}u^{\nu}+P\eta^{\mu\nu}\,,
\end{equation}
and governed by the conservation equation
\begin{equation}
\nabla_{\mu}T^{\mu\nu}=0\,.\label{nablaT}
\end{equation}
Let's denote the temperature of one system by $T_l$ and the other
by $T_r$ and assume that the energies of two systems are
just a function of temperature, i.e.,
$\mathcal{E}_{l,r}=\mathcal{E}(T_{l,r})$. Now we want to bring two systems into contact at $t=0$ and study the possible solutions. Indeed, this is a Riemann problem with
initial values $T_l$ for $x <0$ and $T_r$ for $x > 0$. 
As mentioned in the introduction, this problem has been recently studied for conformal fluid
\cite{Bernard:2013,Bhaseen:2013ypa,Chang:2013gba}. Our goal is to study the problem for an arbitrary equation of state while we follow an approach adopted in \cite{Bhaseen:2013ypa} to examine a possible development of the steady state through the propagation of shock waves emanating from the contact point. 

It is more evident that the
conservation law (\ref{nablaT}) is a system of hyperbolic
equations when it is written as follows
\begin{eqnarray}
\partial_t T^{tt}+\partial_x T^{xt}&=&0\,,\nonumber\\
\partial_t T^{tx}+\partial_x T^{xx}&=&0\,.\label{hyperT}
\end{eqnarray}
Therefore comparing with (\ref{hypersys}) we get
\begin{equation}
Q=\begin{pmatrix}T^{tt}\\T^{tx}\end{pmatrix}\,,\qquad
F=\begin{pmatrix}T^{tx}\\T^{xx}\end{pmatrix}.\label{QF}
\end{equation}
Then all we need to solve for the shock solution are the
Rankine–-Hugoniot jump conditions:
\begin{equation}\label{jumprel}
u_s(Q_l-Q_r)=F_l-F_r
\end{equation}
where $Q$ and $F$ are given by (\ref{QF}) with $Q_l>Q_r$ and
$F_l>F_r$.

The existence and uniqueness of the shock solution for an arbitrary EOS is highly
influenced by the thermodynamic properties of the fluid as well as
the nature of the conservation laws. Nevertheless, for
a relativistic perfect fluid satisfying the hyperbolic laws
(\ref{hyperT}), the existence and uniqueness of the shock solutions has been shown in \cite{Smoller} provided the
difference between quantities on the left and right side of
discontinuity is small and the following relations are fulfilled:
\begin{eqnarray}
c_s<1\,, \qquad\frac{d^2P}{d\mathcal{E}^2}\geqq
-2\frac{(1-c_s^2)c_s^2}{P+\mathcal{E}}\,.\label{constraints}
\end{eqnarray}
The first inequality ensures the physical requirement that the
speed of sound is less than the speed of light, where we have set
$c=1$ throughout this paper.

In the following we study the steady state solutions considering
three types of equation of state: a) linear EOS, i.e., $P=\sigma
\mathcal{E}$, b) the EOS for  a $(d+1)$-dimensional QFT introduced in
\cite{Buchel:2013lla} which is in fact a holographic CFT perturbed by a relevant operator, and c) general barotropic EOS, i.e.,
$P=P(\mathcal{E})$ where we assume the left and right quantity are
slightly different, i.e.,
$\mathcal{E}_l-\mathcal{E}_r\ll\mathcal{E}_{l,r}\,$.

\subsection{Linear EOS}

Consider we start with two sem-infinite fluids described by a
linear EOS:
\begin{equation}
P=\sigma \mathcal{E}\,,\label{linear}
\end{equation}
where from the definition (\ref{soundspeed}) we infer
$\sigma=c_s^2$. Next we bring two systems into instantaneous
thermal contact at $x, t=0$. We show that at long enough time the
steady state forms and we obtain the properties of that. In fact
what we have, looks like a Riemann problem with initial values
\begin{equation}
\mathcal{E}_0(x)=\left\{ \begin{array}{ll}
         \mathcal{E}_l \quad& \mbox{if $x < 0$},\\
        \mathcal{E}_r \quad& \mbox{if $x > 0$}.\end{array} \right.
\end{equation}
A consistent solution has two shock waves propagating in opposite
direction with respect to each other and therefore all we need to
solve are the jump conditions (\ref{jumprel}) for EOS
(\ref{linear}). At long enough time we have three regions:

1. region left described by an ideal fluid with energy momentum
tensor:
\begin{equation}
T^{\mu\nu}_l=\begin{pmatrix}\mathcal{E}_l&0&0&0\\0&\sigma
\mathcal{E}_l&0&0\\\vdots&\vdots&\ddots&\vdots\\0&0&0&\sigma\mathcal{E}_l\end{pmatrix}\label{TmnL}
\end{equation}

2. steady state region in the middle described by a boosted fluid
with energy $\mathcal{E}$ and the boost velocity $v$:
\begin{equation}
T^{\mu\nu}_s=\sigma\mathcal{E}_l\left(\eta^{\mu\nu}+(\sigma^{-1}+1)u^{\mu}u^{\nu}\right)\,,\label{Tmns}
\end{equation}
where $\eta^{\mu\nu}=\textrm{diag}(-1,1,\cdots,1)$ and
$u^{\mu}=\gamma(1,v,0,\cdots,0)$ with $\gamma=1/\sqrt{1-v^2}$.

3. region right described by an ideal fluid with energy momentum
tensor:
\begin{equation}
T^{\mu\nu}_r=\begin{pmatrix}\mathcal{E}_r&0&0&0\\0&\sigma
\mathcal{E}_r&0&0\\\vdots&\vdots&\ddots&\vdots\\0&0&0&\sigma\mathcal{E}_r\end{pmatrix}\label{TmnR}
\end{equation}
Note that $\mathcal{E}_l>\mathcal{E}>\mathcal{E}_r$. Now we apply
the jump relation (\ref{jumprel}) for two shock waves:

I. One shock wave is moving with the speed $u_l$ to the negative
$x$ direction from steady state region with energy $\mathcal{E}$
to the left region with energy $\mathcal{E}_l>\mathcal{E}$,
therefore we have:
\begin{eqnarray}
u_l(Q_l-Q_s)=F_l-F_s\,,\label{uL}
\end{eqnarray}
where $Q$ and $F$ are given in (\ref{QF}).

II. Another shock wave is moving with the speed $u_r$ to the
positive $x$ direction from steady state region with energy
$\mathcal{E}$ to the right region with temperature
$\mathcal{E}_r<\mathcal{E}$, therefore we have:
\begin{eqnarray}
u_r(Q_s-Q_r)&=&F_s-F_r\,.\label{uR}
\end{eqnarray}
combining equations (\ref{uL}) and (\ref{uR}) immediately gives:
\begin{equation}
T^{tx}_s=\frac{T^{xx}_l-T^{xx}_r}{u_{l}+u_{r}}=\sigma\left(\frac{\mathcal{E}_l-\mathcal{E}_r}{u_l+u_r}\right)\,,\label{Ttx}
\end{equation}
where we have used (\ref{TmnL}) and (\ref{TmnR}). Moreover,
substituting (\ref{TmnL}-\ref{Tmns}) into the equations
(\ref{uL}-\ref{uR}), we obtains the four unknowns
$\{u_l,u_r,\mathcal{E},v\}$ in terms of the known boundary
conditions, i.e., $\mathcal{E}_r(x>0)$ and $\mathcal{E}_l(x<0)$
as:
\begin{eqnarray}
&&u_l=\sigma\sqrt{\frac{\bar{\chi}+\sigma^{-1}}{\bar{\chi}+\sigma}}\,,\qquad
u_r=\sqrt{\frac{\bar{\chi}+\sigma}{\bar{\chi}+\sigma^{-1}}}\,,\label{ucon}\\
&&\mathcal{E}=\sqrt{\mathcal{E}_l\mathcal{E}_r}\,,\qquad
v=\frac{\bar{\chi}-1}{\sqrt{(\bar{\chi}+\sigma^{-1})(\bar{\chi}+\sigma)}}\label{vcon}
\end{eqnarray}
where $\bar{\chi}\equiv\sqrt{\mathcal{E}_l/\mathcal{E}_r}$. One
should note that $\sigma=c_s^2<1$ to fulfil relativistic
constraint. Accordingly the second expression in
(\ref{constraints}) is satisfied since $0\leq\sigma<1$ and
therefore the above shock solution is the only solution to this
set up for all $\mathcal{E}_l$ and $\mathcal{E}_r$.\footnote{It is proved that for a linear EOS the uniqueness constraint (\ref{constraints}) is valid for all $\mathcal{E}_l$ and $\mathcal{E}_r$ while for general EOS the constraint is valid provided left and right quantities are close. We refer the interested reader to \cite{Smoller} for the proof.} Moreover, for the special case of $\sigma=1/d$, where $d$ is the spatial dimensions, the
equations (\ref{Ttx}-\ref{vcon}) reduced to the one in
\cite{Bhaseen:2013ypa} for conformal fluid considering the fact
that the energy of the conformal fluid is related to the
temperature as $\mathcal{E}\propto T^{d+1}$ thus $\bar{\chi}$ here
coincide with $\chi$ in that reference.

\subsection{A QFT model}

We start with a $(d+1)$-dimensional QFT in the regions left and right where the
action is given by
\begin{equation}
S_{QFT}=S_{CFT}+\lambda\int d^{d+1}x\, \mathcal{O}(x)\label{qftaction}
\end{equation}
where $\lambda$ has dimension $d+1-\Delta$ and the dimensionless
quantity $\lambda/T^{d+1-\Delta}\ll1$. Note that the unitarity bound for a scalar operator introduces a lower bound on the conformal dimension as $\Delta\ge (d-1)/2$, for most recent study see \cite{Friedan:2015xea}. Also the action (\ref{qftaction}) describes a holographic CFT perturbed by a \textit{relevant operator} which requires $\Delta < d+1$ and in the following we investigate the development of the steady state for such a fluid. Recently in \cite{Bernard:2015aa}, a similar study has been done for a (1+1)-dimensional CFT which is perturbed by an \textit{irrelevant operator}. 

The energy density and pressure of a fluid described by (\ref{qftaction}) at finite temperature has been studied perturbatively in \cite{Buchel:2013lla} and 
is given by:
\begin{eqnarray}
\mathcal{E}(T)=\mathcal{A}T^{d+1}\left(1-\alpha\left(\frac{\lambda}{T^{d+1-\Delta}}\right)^2\right)+\cdots\,,\\
\mathcal{P}(T)=\frac{\mathcal{A}}{d}T^{d+1}\left(1-\left(\frac{\lambda}{T^{d+1-\Delta}}\right)^2\right)+\cdots\,,\label{EPQFT}
\end{eqnarray}
where $\alpha=(2\Delta-d-2)/d$ and $\mathcal{A}$ is proportional
to the central charge of CFT.

Using (\ref{soundspeed}) the speed of sound up to second order in
perturbative expansion is given by
\begin{equation}
c^2_{s,QFT}=c^2_{s,CFT}\left[1+\frac{2(d+1-2\Delta)(d+1-\Delta)}{d(d+1)}\frac{\lambda^2}{T^{2(d+1-\Delta)}}\right]\,,\label{CsQFT}
\end{equation}
where $c^2_{s,CFT}=1/d$. The second term in the square bracket is
negative for $(d+1)/2<\Delta<d+1$ indicating that
$c^2_{s,QFT}<c^2_{s,CFT}$. This is indeed a reasonable
expectation: in the study of QCD thermodynamics in four
dimensions, it is known that in the hot QCD plasma the speed of
sound is approaching the Stefan-Boltzmann limit, equal to
conformal limit, i.e., $c^2_{s,SB}=1/3$, from below
\cite{Borsanyi:2010cj}. For $\Delta<(d+1)/2$ it is
evident that $c^2_{s,QFT}>c^2_{s,CFT}$, 
violating the Stefan-Boltzman limit, yet an allowed solution by unitarity considerations. Also,
using the expression (\ref{CsQFT}) one can compare the speed of
sound in the left and right fluid. For the conformal dimension in
the range $\frac{d+1}{2}\leq\Delta < d+1$ the second term in the
bracket is negative, therefore assuming $T_l>T_r$ one easily
conclude $c^2_{s,QFT,l}>c^2_{s,CFT}>c^2_{s,QFT,r}$ which supports
our intuition that the speed of sound is larger in a fluid with
higher temperature. However, for $\frac{d-1}{2}< \Delta <
\frac{d+1}{2}$ the second term in the bracket in (\ref{CsQFT}) is
positive which yields to $c^2_{s,QFT,l}<c^2_{s,CFT}<c^2_{s,QFT,r}$
while $T_l>T_r$. Although this is a mathematically accepted
solution, it is not physically relevant case.

Moreover, plugging the expressions (\ref{EPQFT}) into equations
(\ref{hyperT}-\ref{jumprel}) we will get the expression for the
temperature of steady state as:
\begin{eqnarray}
T=T_c\left(1+
\frac{\tau\lambda^2}{T_c^{2(d+1-\Delta)}}\right)+\cdots\,,\label{TQFT}
\end{eqnarray}
where $T_c=\sqrt{T_l T_r}$ is the temperature obtained for
conformal case, and
\begin{eqnarray}
\tau=\frac{(1-\chi^{-\delta})}{2d(d+1)(\chi+1)}\left[d(1-\delta)(1-\chi^{\delta+1})+(d-\delta)(1-\chi^{\delta-1})\chi\right]\,,\label{T1}
\end{eqnarray}
where we have defined
\begin{equation}
\chi\equiv\left(\frac{T_l}{T_r}\right)^{\frac{d+1}{2}}\,,\qquad\delta\equiv
2\left(1-\frac{\Delta}{d+1}\right)\,.
\end{equation}
Note that $\chi>1$ since we have assumed $T_l>T_r$. Examining the
expression (\ref{TQFT}) reveals that depending on the value of
conformal dimension $\Delta$ the temperature $T$ of the steady
state starting with the initial perturbed fluids could be smaller
or larger than the temperature $T_c$ of the steady state starting
with the initial conformal fluids. Indeed, $\delta < 1$ for
$\Delta > (d+1)/2$ which result in both terms in square bracket
become negative. That is, in this regime $\tau$ is negative and
therefore $T < T_c$. However, $\delta>1$ for $\Delta < (d+1)/2$
which yields positive $\tau$ and $T>T_c$. More precisely one gets
\begin{eqnarray}
&&T < T_c\qquad \mathrm{for} \qquad \frac{d+1}{2}\leq \Delta < d+1\,,\nonumber\\
&&T>T_c\qquad \mathrm{for} \qquad \frac{d-1}{2}< \Delta <
\frac{d+1}{2}\,.
\end{eqnarray}
The above result is valid independent of the value of $\chi$,
i.e., relative ratio of right and left temperature, however the
temperature $T$ deviates more from $T_c$ when the difference
between right and left temperature increases.

Also one can obtain the boost velocity as:
\begin{equation}
v=v_c\left(1+\frac{\nu\lambda^2}{T_c^{2(d+1-\Delta)}}\right)+\cdots\,,\label{vQFT}
\end{equation}
where $v_c$ is given in equation (\ref{vcon}) for the conformal
fluid by replacing $\bar{\chi}\rightarrow\chi$ and
\begin{equation}
\nu=-\tau\left[\frac{(d+1)^3}{2d}\frac{\chi(\chi+1)(1+\chi^{-\delta})}{(\chi-1)(1-\chi^{-\delta})(\chi+d)(\chi+d^{-1})}\right]\,.\label{v1}
\end{equation}
The square bracket in the above expression is always positive then
the sign of $\nu$ is determined by the sign of $\tau$ which we
already discussed. Therefore from the expression (\ref{vQFT}), one
can deduce that the boost velocity is smaller or larger than the
conformal case depending on the value of conformal dimension
$\Delta$, i.e.,
\begin{eqnarray}
&&v > v_c\qquad \mathrm{for} \qquad \frac{d+1}{2} < \Delta < d+1\,,\nonumber\\
&&v<v_c\qquad \mathrm{for} \qquad \frac{d-1}{2}< \Delta <
\frac{d+1}{2}\,.
\end{eqnarray}
One can also obtain the shock speed in the left and right fluid as
\begin{eqnarray}
&&u_l=u_{c,\,l}\left(1+\frac{\mathcal{U}_l\lambda^2}{T_c^{2(d+1-\Delta)}}\right)+\cdots\,,\nonumber\\
&&u_r=u_{c,\,r}\left(1+\frac{\mathcal{U}_r\lambda^2}{T_c^{2(d+1-\Delta)}}\right)+\cdots\,.\label{uQFT}
\end{eqnarray}
where $u_{c,\,l}$ and $u_{c,\,r}$ are given in equation (\ref{ucon}) for conformal fluid by replacing $\bar{\chi}\rightarrow\chi$ and
\begin{eqnarray}
&&\mathcal{U}_l=\frac{\nu (d-1+2 \chi )}{1+d}+\frac{(1+d)^2
\tau+\alpha(1+d)\chi ^{-\delta}-\alpha
(d+\chi^{-1})+\chi^{-1}-1}{d\chi^{-2}(\chi-1 )
(d^{-1}+ \chi )}\,,\\
&&\mathcal{U}_r=\frac{\nu (d-1 +2 \chi^{-1})}{1 +d }-\frac{(1+d)^2
\tau+\alpha(1+d) \chi ^{\delta}-\alpha (d+\chi)+\chi-1}{(\chi-1 )
(d+\chi )}
\end{eqnarray}
It could be shown
\begin{eqnarray}
&&u_{l,r} < u_{c,\,l,r}\qquad \mathrm{for} \qquad \frac{d+1}{2} < \Delta < d+1\,,\nonumber\\
&&u_{l,r}>u_{c,\,l,r}\qquad \mathrm{for} \qquad \frac{d-1}{2}<
\Delta < \frac{d+1}{2}\,.
\end{eqnarray}
To check if the above solution is the unique solution of the
problem, one needs to check the inequalities (\ref{constraints}).
From equation (\ref{CsQFT}) it is evident that for small
perturbations, i.e., $\lambda/T^{d+1-\Delta}\ll1$, the first
constraint in (\ref{constraints}) is always fulfilled.
The second constraint is always satisfied for all $\mathcal{E}_l$ and $\mathcal{E}_r$ for the conformal fluid and then it holds as well in our QFT model
which is just a small perturbation around CFT. Therefore the shock waves emanating at the contact point are stable and the results obtained for the steady state hold for all $\mathcal{E}_l$ and $\mathcal{E}_r$. 

It might be interesting to compare the results of our QFT model in $(1+1)$-dimensions with that of \cite{Bernard:2015aa}; note that the former is a perturbed CFT by an irrelevant operator while the latter is a CFT perturbed by a relevant operator. Let?s start with the left and right fluid which is described by (\ref{EPQFT}) with $d=1$ and $\Delta=1$, i.e.,
\begin{eqnarray}
\mathcal{E}(T)=\mathcal{A}T^{2}\left(1-\alpha\,\frac{\lambda^2}{T^2}\right)+\cdots\,,\\
\mathcal{P}(T)=\mathcal{A}T^{2}\left(1-\frac{\lambda^2}{T^2}\right)+\cdots\,.\label{EPQFTd-1}
\end{eqnarray}
The temperature and boost velocity of the steady state which develops in the intermediate region would be specified by (\ref{TQFT}) and (\ref{vQFT}), respectively, while the speed of shocks moving to the left and right are given by (\ref{uQFT}). However, one would immediately recover that the subleading terms in all the expansions vanish for $d=1$ and $\Delta=1$ and obtain 
\begin{eqnarray}
T=T_c\,,\qquad v=v_c\,,
\qquad u_l=u_{l,c}\,,\qquad u_r=u_{r,c}\,, \label{Tvud-1}
\end{eqnarray}
for the steady state which is the same with that of conformal fluids. That is, even though the initial fluids that we start with on the left and right are perturbed CFTs given by EOS (\ref{EPQFTd-1}), the steady state developing in between left and right regions shows the same properties as one starts with conformal fluids as the initial states to start with. This is different from the results in \cite{Bernard:2015aa} for a steady state interpolating between left and right CFTs perturbed by a relevant operator where they observe deviation from the conformal case for the temperature, shock speeds and etc.  

We emphasis that expression (\ref{Tvud-1}) is only true for steady state if the conformal dimension of the irrelevant operator is set to one in $(1+1)$ dimensions. For any other allowed values of $\Delta$ one should use expansions (\ref{TQFT}, \ref{vQFT}, \ref{uQFT}) to express the properties of steady state and it definitely deviates from the conformal case. However, similar story happens in an arbitrary dimensions if one chooses an operator where $\Delta=(d+1)/2$.

\subsection{Barotropic fluid}

Consider a barotropic fluid for which pressure is a
function of energy density, i.e.,
\begin{equation}
P=P(\mathcal{E})\,,
\end{equation}
where the energy density is only a function of temperature,
$\mathcal{E}=\mathcal{E}(T)$. We also assume that the difference
between left and right temperature is very small compared to
either left or right temperature which implies
\begin{equation}
\mathcal{E}_l-\mathcal{E}_r\ll\mathcal{E}_{l,r}\,,
\end{equation}
with $\mathcal{E}_l>\mathcal{E}_r$.
Now one can use the Taylor expansion for pressure around some
$\mathcal{E}_0$ as
\begin{equation}
P(\mathcal{E})=P_0+c_s^2(\mathcal{E}-\mathcal{E}_0)+\frac{1}{2}\kappa(\mathcal{E}-\mathcal{E}_0)^2+\cdots
\end{equation}
where we have defined $P(\mathcal{E}_0)\equiv P_0$ and
\begin{equation}
\kappa\equiv\frac{d^2P}{d\mathcal{E}^2}\,,
\end{equation}
is a characteristic parameter of the fluid and should satisfy the second inequality in
(\ref{constraints}), in order to have a unique shock solution
to the Riemann problem with initial values $\mathcal{E}_l$ for
$x<0$ and $\mathcal{E}_r$ for $x>0$. Now solving equations
(\ref{hyperT}-\ref{jumprel}) perturbatively we will get the
following expression for the energy density of the steady state
fluid
\begin{equation}
\mathcal{E}=\mathcal{E}_0+\frac{1}{16 c_s^2}\left(\kappa
-\frac{2(1+c_s^2)c_s^2}{H_0}\right)\Delta\mathcal{E}^2+\cdots\,,\label{Edensity}
\end{equation}
where we have defined $H_0\equiv P_0+\mathcal{E}_0$ which is the
enthalpy of the equilibrium state where
$\mathcal{E}_l=\mathcal{E}_r$. Also, we have chosen
$\mathcal{E}_0$ to be the mean energy of the left and right fluid,
i.e.,
\begin{equation}
\mathcal{E}_0=\frac{\mathcal{E}_l+\mathcal{E}_r}{2}\,,
\end{equation}
and
\begin{equation}
\Delta\mathcal{E}=\mathcal{E}_l-\mathcal{E}_r\,.
\end{equation}
Note that the first correction which appears in energy density is
of second order. Indeed, this is not surprising as one would
expect only even powers of $\Delta\mathcal{E}$ appears in the
energy expansion simply because the energy of the steady state
should not change if we start by the initial condition where we
have replaced $\mathcal{E}_l\leftrightarrow\mathcal{E}_r$.
Furthermore, the boost velocity is given by
\begin{equation}
v=\frac{c_s}{2\,H_0}\,\Delta\mathcal{E}+\cdots\,.\label{boostV}
\end{equation}
Here, we expect only odd powers of $\Delta\mathcal{E}$ appears in
the boost expansion since again the magnitude of the boost
velocity should not change by replacing
$\mathcal{E}_l\leftrightarrow\mathcal{E}_r$, however one need to
boost in the opposite direction.

The speed of sound in the left and right fluid to the first order
in $\Delta\mathcal{E}$ is given by
\begin{eqnarray}
&&c_{s,l}=c_s(1+\frac{\kappa}{4c_s^2}\Delta\mathcal{E}+\cdots)\,,\nonumber\\
&&c_{s,r}=c_s(1-\frac{\kappa}{4c_s^2}\Delta\mathcal{E}+\cdots)\,.\label{CsLR}
\end{eqnarray}
We did not include the second order term in the above expansion in
order to avoid introducing a new characteristic parameter for the
fluid. For $\kappa>0$ it is clear that $c_{s,l}>c_s>c_{s,r}$ which
is consistent with our assumption for the temperature that
$T_l>T_r$. On the other hand for $\kappa<0$ we get
$c_{s,r}>c_s>c_{s,l}$ while we have assumed $T_l>T_r$. This is not
what we usually expect in a normal physical system. Note that the
speed of sound on left and right would be replaced under
$\mathcal{E}_l\leftrightarrow\mathcal{E}_r$.

We can further obtain the speed of shock waves on the left and
right fluid as
\begin{eqnarray}
&&u_l=c_{s,\,l}\left[1-\frac{1}{8c^2_{s,\,l}}\left(\kappa+2\frac{(1-c_{s,l}^2)c_{s,l}^2}{H_0}\right)\Delta\mathcal{E}+\cdots\right]\,,\nonumber\\
&&u_r=c_{s,\,r}\left[1+\frac{1}{8c_{s,\,r}^2}\left(\kappa+2\frac{(1-c_{s,\,r}^2)c^2_{s,\,r}}{H_0}\right)\Delta\mathcal{E}+\cdots\right]\,.\label{uLuR}
\end{eqnarray}
The bracket in the second term is always positive due to the
stability constraint (\ref{constraints}), therefore the left shock
is subsonic compared to the speed of sound on the left while the
right shock is supersonic with respect to the speed of sound on
the right.

If the characteristic parameter $\kappa>0$ then the second
constraint in (\ref{constraints}) is always satisfied and
therefore the above solution is a unique solution. This is indeed a
well-known fact, even for a classical fluid, that $\kappa>0$ is a
sufficient condition for the existence and uniqueness of shock
waves to the Riemann problem, known as Bethe-Weyl theorem
\cite{Bethe,Weyl}. However for an arbitrary equation of state with $\kappa<0$, it is
still possible to have a unique shock solution for the Riemann
problem as long as the inequality in (\ref{constraints})
fulfilled.

\section{Discussion} \label{conclusion}

We argued the possibility of performing the NESS when we bring two copies of systems at different temperatures into a thermal contact in the framework of Bernard and Doyon \cite{Bernard:2013}, however, our systems enjoy more general EOS than the conformal fluid. The key feature in the BD set-up is that the steady state is \textit{not} driven by the temperature gradient. There are no external reservoirs, rather the initial systems play the role of heat baths at sufficiently large scales. Furthermore, we used the holographic insight of \cite{Bhaseen:2013ypa} to describe the steady state as an intermediate Lorentz boosted state of the initial systems, interpolating between two asymptotic heat baths after long enough time. In this approach the formation of the steady state relies on the two single shock waves emanating at the point of contact moving in the opposite directions. Since both systems are assumed to be semi-infinite then the claim is that there is no chance for the shocks to reflect back, forming a cascade and thermalizing the system. Although this seems to be sufficient condition to construct the steady state in the case of CFT, it is not enough for a system with a general EOS, e.g., barotropic fluid that we studied in this paper. In general, in order to perform the steady state, shock waves should not split otherwise thermalization may happen. This is due to the fact that if they split, they may move in the opposite directions and form composite waves leading to cascades and finally thermalizing the system. For a relativistic perfect fluid with general EOS the condition (\ref{constraints}) is sufficient in order to avoid splitting and to have stable shock solutions. Therefore the development of a steady state is guaranteed and one can obtain the properties of this NESS in terms of the characteristics of the fluid and the initial values, see equations (\ref{Edensity}) and (\ref{boostV}).

Nonetheless, in our studies we only considered perfect fluids, for which viscosity is zero. It is interesting to see how the construction of the steady states would be affected in this set-up, if we move away from ideal hydro and take into account the viscosity. Since performing of the steady state depends on the propagation of shock waves as mentioned before, then we can turn around and ask how viscosity will affect the formation of shock waves at the point of contact. More precisely, one should investigate the possibility of shock solutions for the initial condition Riemann problem in viscous fluid. In fact, this question has been already studied in literature, e.g., see \cite{Bouras:2009nn} where the solutions to the relativistic Riemann problem for viscous fluid has been investigated numerically. As a result, by varying the ratio of shear viscosity to entropy density, i.e., $\eta/s$, from zero to infinity, a transition from ideal shock waves to viscous one has been shown; starting from ideal fluid with zero viscosity one obtains shock waves with zero width. By increasing $\eta/s$ the solutions with non-zero width will appear, the so-called viscous shocks. An upper limit for the $\eta/s$ has been estimated for which shocks can still be observed experimentally on the proper time scale. However, as one continues to increase the ratio $\eta/s$ above this upper limit, the free-streaming will occur, i.e., shock solution is completely washed out.

\section*{Acknowledgment}

I would like to express my gratitude to Rob Myers for the initial motivation and for his continuous guidance throughout the progress of this paper. I would also like to thank Perimeter Institute for hospitality at an early stage of this work. The research of R.P. is supported by Icelandic Research Fund grant 130131-053.

\bibliography{shock1}{}

\providecommand{\href}[2]{#2}\begingroup\raggedright\begin{thebibliography}{10}

\bibitem{Bernard:2013}
D.~Bernard and B.~Doyon, {\it {Non-equilibrium steady-states in conformal field
  theory}},  \href{http://xxx.lanl.gov/abs/1302.3125}{{\tt arXiv:1302.3125}}.

\bibitem{Chang:2013gba}
H.-C. Chang, A.~Karch, and A.~Yarom, {\it {An ansatz for one dimensional steady
  state configurations}},  {\em J.Stat.Mech.} {\bf 06} (2014) 018,
  [\href{http://xxx.lanl.gov/abs/1311.2590}{{\tt arXiv:1311.2590}}].

\bibitem{Bhaseen:2013ypa}
M.~Bhaseen, B.~Doyon, A.~Lucas, and K.~Schalm, {\it {Far from equilibrium
  energy flow in quantum critical systems}},
  \href{http://xxx.lanl.gov/abs/1311.3655}{{\tt arXiv:1311.3655}}.

\bibitem{Lax:1957}
P.~D. Lax, {\it Hyperbolic systems of conservation laws},  {\em II. Commun.
  Pure Appl. Math.} {\bf 10} (1957) 537--566.

\bibitem{Lax:1972}
P.~D. Lax, {\it The formation and decay of shock waves},  {\em The American
  Mathematical Monthly} {\bf 79} (1972), no.~3 pp. 227--241.

\bibitem{Menikoff:1989ka}
R.~Menikoff and B.~J. Plohr, {\it {The Riemann problem for fluid flow of real
  materials}},  {\em Rev. Mod. Phys.} {\bf 61} (1989) 75--130.

\bibitem{Smoller}
J.~Smoller and B.~Temple, {\it Global solutions of the relativistic euler
  equations},  {\em Communications in Mathematical Physics} {\bf 156} (1993)
  67--99.

\bibitem{FLM:357096}
J.~M. Mart{\'\i} and E.~M{\"u}ller, {\it The analytical solution of the riemann
  problem in relativistic hydrodynamics},  {\em Journal of Fluid Mechanics}
  {\bf 258} (1, 1994) 317--333.

\bibitem{Buchel:2013lla}
A.~Buchel, L.~Lehner, R.~C. Myers, and A.~van Niekerk, {\it {Quantum quenches
  of holographic plasmas}},  {\em JHEP} {\bf 1305} (2013) 067,
  [\href{http://xxx.lanl.gov/abs/1302.2924}{{\tt arXiv:1302.2924}}].

\bibitem{Bressan}
A.~Bressan, {\it Hyperbolic conservation laws},  in {\em Mathematics of
  Complexity and Dynamical Systems} (R.~A. Meyers, ed.), pp.~729--739.
\newblock Springer New York, 2011.

\bibitem{Bethe}
H.~Bethe, {\it On the theory of shock waves for an arbitrary equation of
  state},  in {\em Classic Papers in Shock Compression Science} (J.~Johnson and
  R.~Ch{\'e}ret, eds.), High-Pressure Shock Compression of Condensed Matter,
  pp.~421--495.
\newblock Springer New York, 1998.

\bibitem{Zeldovich}
J.~Zeldovich, {\it On the possibility of rarefaction shock waves},  {\em Zh.
  Eksp. Teor. Fiz.} {\bf 4} (1946) 363--364.

\bibitem{Thompson}
P.~A. Thompson, {\it A fundamental derivative in gas dynamics},  {\em Phys.
  Fluids} {\bf 14(9)} (1971) 1843--1849.

\bibitem{Friedan:2015xea}
D.~Friedan and C.~A. Keller, {\it {Cauchy conformal fields in dimensions
  $d>2$}},  \href{http://xxx.lanl.gov/abs/1509.0747}{{\tt arXiv:1509.0747}}.

\bibitem{Bernard:2015aa}
D.~Bernard and B.~Doyon, {\it A hydrodynamic approach to non-equilibrium
  conformal field theories},  \href{http://xxx.lanl.gov/abs/1507.0747}{{\tt
  arXiv:1507.0747}}.

\bibitem{Borsanyi:2010cj}
S.~Borsanyi, G.~Endrodi, Z.~Fodor, A.~Jakovac, S.~D. Katz, et~al., {\it {The
  QCD equation of state with dynamical quarks}},  {\em JHEP} {\bf 1011} (2010)
  077, [\href{http://xxx.lanl.gov/abs/1007.2580}{{\tt arXiv:1007.2580}}].

\bibitem{Weyl}
H.~Weyl, {\it Shock waves in arbitrary fluids},  {\em Communications on Pure
  and Applied Mathematics} {\bf 2} (1949), no.~2-3 103--122.

\bibitem{Bouras:2009nn}
I.~Bouras, E.~Molnar, H.~Niemi, Z.~Xu, A.~El, et~al., {\it {Relativistic shock
  waves in viscous gluon matter}},  {\em Phys.Rev.Lett.} {\bf 103} (2009)
  032301, [\href{http://xxx.lanl.gov/abs/0902.1927}{{\tt arXiv:0902.1927}}].

\end{thebibliography}\endgroup
\bibliographystyle{JHEP}

\end{document}